# Universality of critical behaviors in the three-dimensional (3D) Ising magnets

Zhidong Zhang

Shenyang National Laboratory for Materials Science, Institute of Metal Research,

Chinese Academy of Sciences, 72 Wenhua Road, Shenyang, 110016, P.R. China

**Abstract**

This article gives a brief overview on recent advances in experiments of critical exponents in three groups of magnetic materials. Revisiting experimental data verifies that a universality class with the critical exponents $\beta = 3/8$, $\gamma = 5/4$ and $\delta = 13/3$ occurs in the three-dimensional (3D) Ising magnets, such as transition-metal intermetallics, rare-earth transition-metal compounds and manganites. The experimental results agree well with the exact solutions of the 3D Ising models. Furthermore, the topological contributions to critical behaviors in the 3D Ising model are estimated by the difference between the exact solutions and the approximation values.

The corresponding author: Z.D. Zhang, e-mail address: zdzhang@imr.ac.cn

## 1.Introduction

Phase transitions occur in almost every matter in nature, and critical phenomena at/near a critical point of a second-order phase transition are particularly interesting topics in physics. The study of the critical phenomena in many-body interacting spin (or particle) systems is very important for understanding condensed matters (such as, magnets, superconductors, superfluid, etc.) and other physical systems (such as particle physics and high-energy physics). Some prototype spin models, for instance, the Ising model [1] and the Heisenberg model [2], have been invented to describe the many-body interacting spin systems. It is easy to appreciate that the exact solutions of these models provide a full description of the many-body interacting systems. However, solving exactly the many-body interacting models meets three main obstacles: Nonlocality, nonlinearity and noncommutative of operators exist in the transfer matrices of quantum statistical mechanics.

The Ising model is one of the simplest spin models describing the physical systems with many-body interactions [3,4]. The exact solution of the three-dimensional (3D) Ising model is a well-known long-standing problem in physics. In order to solve this hard problem, the present author proposed two conjectures in [5], investigated its mathematical structure in [6], then proved rigorously the two conjectures by a Clifford algebraic approach in collaboration with Suzuki and March [7], and further by a method of Riemann-Hilbert problem in collaboration with Suzuki [8,9]. The critical exponents were derived exactly to be $\alpha = 0$, $\beta = 3/8$, $\gamma = 5/4$, $\delta = 13/3$, $\eta = 1/8$ and $\nu = 2/3$ [5]. Furthermore, the exact solutions of the two-dimensional (2D) Ising model with a transverse field [10] and the 3D $Z_2$ lattice gauge theory [11] were derived by the mappings between these models. Based on these results, topological quantum statistical mechanics and topological quantum field theories were investigated systematically [12]. In addition, the lower bound of computational complexity of several NP-complete problems, such as spin-glass 3D Ising models [13], Boolean satisfiability problems [14], knapsack problems [15] and traveling salesman problems [16], were determined.

Experiments serve as one of the most important standards for judging the validity of a solution and/or its application range. Zhang and March summarized experimental data for the critical exponents in some magnetic materials and at fluid-fluid phase transitions and demonstrated that the 3D Ising universality exists for the critical indices in a certain class of magnets and at fluid-fluid phase transition [17]. In particular, we mention here some typical systems, for instance, Ni ($\beta = 0.373$, $\gamma = 1.28$, $\delta = 4.44$) [18,19], $CrBr_3$ ($\beta = 0.368$, $\gamma = 1.215$, $\delta = 4.31$) [20,21], $LaMn_{0.9}Ti_{01}O_3$ ($\beta = 0.375$, $\gamma = 1.25$, $\delta = 4.11$) [22], $CO_2$ ($\beta = 0.350$, $\gamma = 1.26$, $\delta = 4.60$) [21], Xe ($\beta = 0.350$, $\gamma = 1.26$, $\delta = 4.60$) [21], He ($\beta = 0.359$, $\gamma = 1.24$, $\delta = 4.45$) [21]. After publication of [17], new advances appear, but the data scatter in literature. It is worth collecting the experimental data in the literature to give an overview on the universality of critical behaviors in the 3D Ising magnets.

This article gives an overview on recent advances in the 3D Ising critical behaviors. In Section 2, we summarize experimental results for the critical exponents in the 3D Ising magnets. In Section 3, we discuss briefly the contributions of nontrivial topological structures to the critical behaviors in the 3D Ising model. Section 4 is for conclusions.

## 2. Experimental data for the critical exponents in the 3D Ising magnets

In this section, we give a brief overview on recent advances in experiments for the critical exponents in three groups of magnetic materials, such as transition-metal intermetallics, rare-earth transition-metal compounds and manganites.

### 2.1 Transition-metal intermetallics

Zhang et al. determined the critical exponents of a transition-metal intermetallic $CuCr_2Se$ to be $\beta = 0.372$, $\gamma = 1.277$ and $\delta = 4.749$ [23]. The critical exponents obtained by Li et al. for $CuCr_2Te_4$ are $\beta = 0.369$, $\gamma = 1.27$ and $\delta = 4.7$ [24]. Rduch et al. studied the influence of Ce substitution on the critical properties of $Cd_xCe_yCr_2Se_4$ ferromagnets, and the results for $Cd_{0.96}Ce_{0.03}Cr_2Se_4$ are $\beta = 0.359$, $\gamma = 1.186$ and $\delta = 4.405$ [25], while for $Cd_{0.84}Ce_{0.13}Cr_2Se_4$, $\beta = 0.382$, $\gamma = 1.286$ and $\delta = 4.487$. Rduch et al. also investigated the critical behaviors of the 3D Ising ferromagnets $Cd[Cr_xTi_y]Se_4$ and obtained the critical exponents $\beta = 0.34$, $\gamma = 1.29$ and $\delta = 4.2$ for $CdCr_2Se_4$ [26]. Liu et al. reported the critical exponents of the van der Waals bonded ferromagnet $Fe_{3-x}GeTe_2$, which are $\beta = 0.372$, $\gamma = 1.265$ and $\delta = 4.401$ [27]. Mao et al. focused on the critical properties of the quasi-2D metallic ferromagnet $Fe_{2.85}GeTe_2$, and derived the critical exponents $\beta = 0.361$, $\gamma = 1.225$ and $\delta = 4.382$ [28]. Zhang et al. demonstrated the critical exponents of the quasi-2D ferromagnet $Cr_4Te_5$ to be $\beta = 0.388$, $\gamma = 1.290$ and $\delta = 4.32$ [29]. Purwar et al. studied 3D-Ising-type magnetic interactions and determined the critical exponents $\beta = 0.360$, $\gamma = 1.221$, $\delta = 4.392$ for layered ferromagnetic $Cr_2Te_3$ [30], which can be compared with Wang et al.'s results (β = 0.340, γ = 1.114, δ = 4.276) [31]. It should be noticed that these layered or quasi-2D ferromagnets still behave as 3D-Ising-type magnets, because of their 3D bulk characters. Murugan et al. found the critical exponents β = 0.380, γ = 1.293 and δ = 4.389 in all *d*-metal Heusler alloy $Fe_{30}Cr_{45}V_{25}$ [32].

2.2 Rare-earth transition-metal compounds

Zheng et al. investigated the critical behavior of amorphous $(Gd_4Co_3)_{1-x}Si_x$ alloys and obtained the critical exponents β = 0.359, γ = 1.223 and δ = 4.405 for $(Gd_4Co_3)_{0.95}Si_{0.05}$ [33]. Opletal et al. reported universality classes of isostructural UTX compounds (T = Rh, Co, $Co_{0.98}R_{u0.02}$; X = Ga, Al): URhGa, β = 0.39, γ = 1.19; UCoGa, β = 0.37, γ = 1.26, δ = 4.32; $UCo_{0.98}Ru_{0.02}Al$, β = 0.36, γ = 1.26, δ = 4.5 [34]. Paul-Boncour determined the critical exponents β = 0.358, γ = 1.20 and δ = 4.3 in $Y_{0.9}Pr_{0.1}Fe_2D_{3.5}$ deuteride [35]. Jaballah reported the critical exponents β = 0.362, γ = 1.345 and δ = 4.71 in the nanocrystalline $Pr_2Fe_{16}Al$ [36] and investigated the critical behaviors in cobalt-substituted $Ce_2Fe_{17}$ compound and determined the critical exponents of β = 0.379, γ = 1.17, δ = 4.09 for $Ce_2Fe_{16.4}Co_{0.6}$ [37].

2.3 Manganites

Ghosh et al. determined experimentally the critical exponents in the double-exchange ferromagnet $La_{0.7}Sr_{0.3}MnO_3$ to be β = 0.37, γ = 1.22 and δ = 4.25 [38]. Yang and Lee investigated the critical behaviors in Ti-doped manganites $LaMn_{1-x}Ti_xO_3$ (0.05 ≤ x ≤ 0.2) and found that the critical exponents are very close to the exact solution of the 3D Ising model [22]. Phan et al. reported the critical exponents of $La_{0.7}Ca_{0.3-x}Sr_xMnO_3$ (x = 0, 0.05, 0.1, 0.2, 0.25) single crystals, and for $La_{0.7}Ca_{0.1}Sr_{0.2}MnO_3$, the critical exponents are β = 0.36, γ = 1.22 and δ = 4.4 [39]. Dhahri et al. [40], Omri et al. [41], Dhahri et al. [42], Tka et al. [43] and Ghodhbane et al. [44] studied the critical behaviors in Co-, Ga-, Ti-, Al- and Fe-doped manganites respectively and the critical exponents in these system agree well with the exact solution of the 3DIsing model. More results for the critical behaviors in various element-doped manganites can be found in literatures, for instance, Zhang et al. [45], Mnefgui et al. [46], Phan et al. [47], Dhahri et al. [48], Mahjoub et al. [49,50], Ho et al. [51], Omrani et al. [52], Kumar et al. [53] and Mtiraoui et al. [54].

All the experimental data obtained in above magnetic materials are collected in Table 1. It can be seen from Table 1 that three critical exponents β, γ and δ determined experimentally from magnetization in almost all the magnets agree well with the exact solutions of the 3D Ising model. However, in few cases [48,52,53], only two critical exponents γ and δ fit well with the theoretical ones, while the critical exponent β has a large deviation. It is a fact that only two critical exponents are independent parameters among all the six critical exponents. If we used the two critical exponents γ and δ experimentally determined to calculate the critical exponent β, the calculated value would be close to the exact solution.

Table 1. Experimental data reported for the critical exponents in the 3D Ising magnetic materials.

| Magnetic materials | β | γ | δ | References |
|---|---|---|---|---|
| $CuCr_2Se$ | 0.372 | 1.277 | 4.749 | [23] |
| $CuCr_2Te_4$ | 0.369 | 1.27 | 4.73 | [24] |
| $Cd_{0.96}Ce_{0.03}Cr_2Se_4$ | 0.359 | 1.186 | 4.405 | [25] |
| $Cd_{0.84}Ce_{0.13}Cr_2Se_4$ | 0.382 | 1.286 | 4.487 | [25] |
| $CdCr_2Se_4$ | 0.34 | 1.29 | 4.2 | [26] |
| $Fe_{3-x}GeTe_2$ | 0.372 | 1.265 | 4.401 | [27] |
| $Fe_{2.85}GeTe_2$ | 0.361 | 1.225 | 4.382 | [28] |
| $Cr_4Te_5$ | 0.388 | 1.290 | 4.32 | [29] |
| $Cr_2Te_3$ | 0.360 | 1.221 | 4.392 | [30] |
| $Cr_2Te_3$ | 0.340 | 1.114 | 4.276 | [31] |
| $Fe_{30}Cr_{45}V_{25}$ | 0.380 | 1.293 | 4.389 | [32] |
| $(Gd_4Co_3)_{0.95}Si_{0.05}$ | 0.359 | 1.223 | 4.405 | [33] |
| URhGa | 0.39 | 1.19 | | [34] |
| UCoGa | 0.37 | 1.26 | 4.32 | [34] |
| $UCo_{0.98}Ru_{0.02}Al$ | 0.36 | 1.26 | 4.5 | [34] |
| $Y_{0.9}Pr_{0.1}Fe_2D_{3.5}$ | 0.358 | 1.20 | 4.3 | [35] |
| $Pr_2Fe_{16}Al$ | 0.362 | 1.345 | 4.71 | [36] |
| $Ce_2Fe_{16.4}Co_{0.6}$ | 0.379 | 1.17 | 4.09 | [37] |
| $La_{0.7}Sr_{0.3}MnO_3$ | 0.37 | 1.22 | 4.25 | [38] |
| $LaMn_{0.95}Ti_{0.05}O_3$ | 0.378 | 1.29 | 4.19 | [22] |
| $LaMn_{0.9}Ti_{01}O_3$ | 0.375 | 1.25 | 4.11 | [22] |
| $LaMn_{0.85}Ti_{0.15}O_3$ | 0.376 | 1.24 | 4.16 | [22] |

| $LaMn_{0.8}Ti_{0.2}O_3$ | 0.359 | 1.28 | 4.21 | [22] |
| --- | --- | --- | --- | --- |
| $La_{0.7}Ca_{0.1}Sr_{0.2}MnO_3$ | 0.36 | 1.22 | 4.4 | [39] |
| $La_{0.67}Pb_{0.33}MnO_3$, | 0.367 | 1.22 | 4.32 | [40] |
| $La_{0.75}Ca_{0.08}Sr_{0.17}Mn_{0.95}Ga_{0.05}O_3$ | 0.389 | 1.251 | 4.22 | [41] |
| $La_{0.57}Nd_{0.1}Pb_{0.33}MnO_3$ | 0.371 | 1.380 | 4.270 | [42] |
| $La_{0.57}Nd_{0.1}Pb_{0.33}Mn_{0.95}Ti_{0.05}O_3$ | 0.391 | 1.276 | 4.470 | [42] |
| $La_{0.57}Nd_{0.1}Sr_{0.33}MnO_3$ | 0.366 | 1.265 | 4.23 | [43] |
| $La_{0.57}Nd_{0.1}Sr_{0.33}Mn_{0.95}Al_{0.05}O_3$ | 0.358 | 1.312 | 4.19 | [43] |
| $La_{0.57}Nd_{0.1}Sr_{0.33}Mn_{0.90}Al_{0.10}O_3$ | 0.353 | 1.333 | 4.13 | [43] |
| $La_{0.8}Ba_{0.2}Mn_{0.85}Fe_{0.15}O_3$ | 0.370 | 1.359 | 4.40. | [44] |
| $La_{0.8}Ba_{0.2}Mn_{0.8}Fe_{0.2}O_3$ | 0.365 | 1.227 | 4.36 | [44] |
| $La_{0.8}Ca_{0.2}MnO_3$ | 0.349 | 1.231 | 4.524. | [45] |
| $La_{0.57}Nd_{0.1}Sr_{0.33}MnO_3$ | 0.368 | 1.191 | 4.236 | [46] |
| $La_{0.7}Sr_{0.3}MnO_3$ | 0.387 | 1.166 | 4.01 | [47] |
| $La_{0.7}Ca_{0.2}Sr_{0.1}Mn_{0.85}Cr_{0.15}O_3$ | 0.323 | 1.22 | 4.415 | [48] |
| $Pr_{0.6}Ca_{0.1}Sr_{0.3}Mn_{0.975}Fe_{0.025}O_3$ | 0.370 | 1.22 | 4.29 | [49] |
| $Pr_{0.6}Ca_{0.1}Sr_{0.3}Mn_{0.95}Fe_{0.05}O_3$ | 0.373 | 1.269 | 4.40 | [50] |
| $Pr_{0.6}Ca_{0.1}Sr_{0.3}Mn_{0.925}Fe_{0.075}O_3$ | 0.377 | 1.295 | 4.30 | [50] |
| $La_{0.7}Ba_{0.3}Mn_{0.95}Ti_{0.05}O_3$ | 0.374 | 1.228 | 4.26 | [51] |
| $Pr_{0.6}Ca_{0.1}Sr_{0.3}MnO_3$ | 0.335 | 1.218 | 4.347 | [52] |
| $Pr_{0.58}Er_{0.02}Ca_{0.1}Sr_{0.3}MnO_3$ | 0.336 | 1.177 | 4.216 | [52] |
| $Pr_{0.54}Er_{0.06}Ca_{0.1}Sr_{0.4}MnO_3$ | 0.395 | 1.289 | 4.263 | [52] |

| $La_{0.7}Sr_{0.3}Si_{0.05}Mn_{0.95}O_3$ | 0.94 | 1.20 | 4.4 | [53] |
|---|---|---|---|---|
| $La_{0.65}Sr_{0.2}K_{0.15}MnO_3$ | 0.39 | 1.21 | 4.10 | [54] |

All the experimental data collected in Table 1 and also Table 1 in [17] are illustrated in Figure 1, which shows clearly that the 3D Ising universality forms in these materials with small error ranges. It is clear that the experimental results are consistent with the exact solutions of the 3D Ising models, obtained in [5].

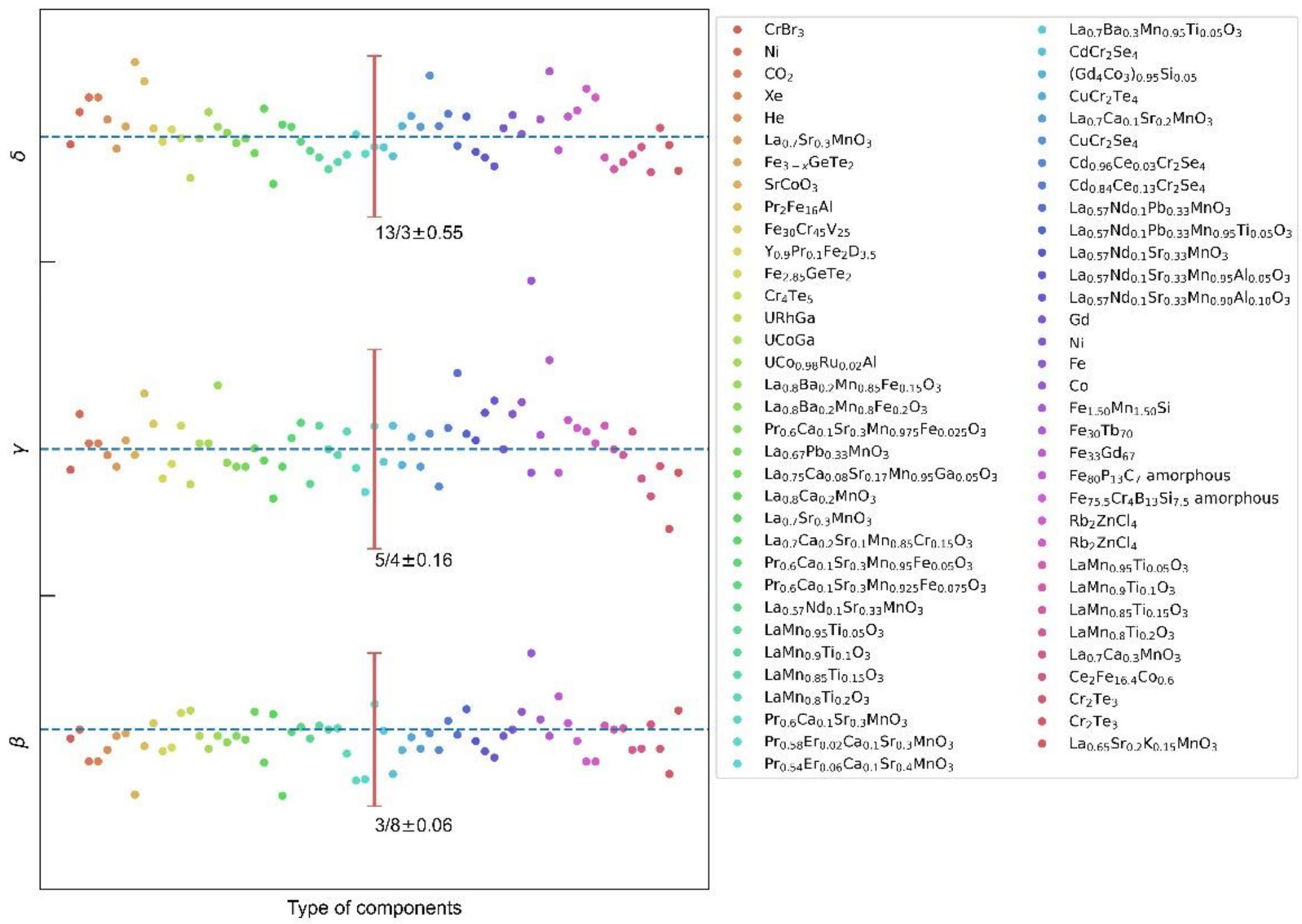

Figure 1. The critical exponents β, γ and δ in some magnetic materials, which show the 3D Ising universality. The experimental data are collected in Table 1 and also Table 1 in [17].

## 3. Contributions of nontrivial topological structures to critical behaviors

One of the important discoveries in our previous work [5-9] is to reveal the existence of nontrivial topological structures (i.e, long-range spin entanglements) in the 3D Ising model, which contribute to thermodynamic properties and critical behaviors. The nontrivial topological structures originate from the contradictory of the 2D character of transfer matrices in quantum statistic mechanism and the 3D character of the spin arrangements in a 3D lattice [11,15]. The topological contributions are achieved by an additional rotation in the (3+1) framework, which represents a Lorentz transformation and also a gauge transformation while generates topological phases in the quaternionic eigenvectors for the many-body interacting systems [5-9]. Since publication of the two conjectures [5], there were the ongoing debates in the community regarding 3D Ising solvability [55-61]. The arguments were mainly focused on disagreement with approximation methods such as conventional low- and high-temperature expansions, Monte Carlo simulations, renormalization group theory, etc. Up to now, the approximations for the critical exponents of the 3D Ising model, which are well-accepted by the community, are $\alpha = 0.109$, $\beta = 0.325$, $\gamma = 1.241$, $\delta = 4.82$, $\eta = 0.031$ and $\nu = 0.630$ [62-64]. Here, we just give an explanation why the multitude of separate determinations of the critical exponents throughout the years, by various independent scientists and using seems completely different techniques coincide. Superficially, all these different techniques (widely accepted by the community) in the deeper level are connected closely. The systematical errors exist seriously in these approximation techniques, which are related directly to the physical conceptions/pictures at the first beginning, neglecting the contributions of the nontrivial topological structures to the partition function, the free energy and the subsequent thermodynamic properties. The systematical errors are intrinsic, which cannot be removed by the efforts of improving technically the precision of these approximation/perturbation techniques. For detailed discussion on the disadvantages of several widely used techniques, readers refer to [5,6,11,65,66]:

The approximation methods, such as the renormalization group theory and Monte Carlo simulations [67,68], are still powerful techniques for the study of the critical phenomena. On one hand, as suggested in [66], one can obtain the topological contributions to the partition function and also the thermodynamic properties by comparing the approximations with the exact solutions. Thus, the nonlocal part of the physical properties (such as spontaneous magnetization) of the 3D Ising model can be obtained by extracting the approximation results from the exact solutions. Here, we define the topological contributions to the critical exponents as $C_I^T$, with $C$ denotes the critical exponents ($\alpha$, $\beta$, $\gamma$, $\delta$, $\eta$, $\nu$), the subscript $I$ denotes the 3D Ising model and the superscript $T$ denotes the topological part. Then we have $C_I^T = C_I^E - C_I^A$, with the superscripts $E$ and $A$ representing the exact solutions and the approximation values. Table 2 shows the exact solutions $C_I^E$, the approximations $C_I^A$ and the topological parts $C_I^T$ of the critical exponents of the 3D Ising model. It is worth noticing that in Table 2, the approximation value for the critical exponent $\gamma$ is consistent with the exact value. This suggest that the nontrivial topological structures contribute less during the simulations of the critical exponent $\gamma$ for magnetic susceptibility $\chi$. The approximation methods for the critical exponents ($\alpha$, $\beta$, $\delta$, $\eta$, $\nu$) have a large deviation with respect to the exact solutions, which should be amended accordingly. On the other hand, one can still use these approximation techniques [69,70], but focus on the structures illustrated in Figure 5 of ref. [9] (also Figure 1 in ref. [69]), which consist of two parts of contributions (local spin alignments and nonlocal long-range spin entanglements). The results obtained by the Monte Carlo method for such structures (including the nonlocal effects) would be close to the exact solutions.

Table 2. The exact solutions $C_I^E$, the approximations $C_I^A$ and the topological parts $C_I^T$ of the critical exponents of the 3D Ising magnets, with $C_I^T = C_I^E - C_I^A$. The approximation values $C_I^A$ are summarized from [62-64], while the exact solutions $C_I^E$ are taken from [5].

| | α | β | γ | δ | η | ν |
|---|---|---|---|---|---|---|
| Exact solution $C_I^E$ | 0 | 3/8 | 5/4 | 13/3 | 1/8 | 2/3 |
| Approximation $C_I^A$ | 0.109 | 0.325 | 1.241 | 4.82 | 0.031 | 0.630 |
| Topologic part $C_I^T$ | -0.109 | 0.050 | 0.009 | -0.487 | 0.094 | 0.036 |

It is worth noticing that the nontrivial topological structures contribute less during the simulations of the critical exponent γ for magnetic susceptibility, while the other exponents have a large deviation with respect to the exact solutions. The reasons are interpreted as follows: In general, all the critical exponents describing the magnetic systems are affected by the nontrivial topological structures, but the effects may be in different levels. The critical exponents β and δ depend on magnetization $M = -\frac{\partial f}{\partial H}$, which is the first derivative of the free energy *f* with respect to the magnetic field *H*. The critical exponent γ for magnetic susceptibility $\chi = \frac{\partial M}{\partial H}$ relies on the second derivative of the free energy *f* with respect to the magnetic field *H*, namely, $\chi = -\frac{\partial^2 f}{\partial H^2}$. Thus, the contributions of the nontrivial topological structures to the critical exponent γ are less than those for the critical exponents β and δ. On the other hands, the critical exponents η and ν depend on the correlation function $\Gamma_c(\mathrm{r})$ and the correlation length $\xi = \frac{1}{\kappa_x}$ ($\kappa_x$ is the true range of the correlation), respectively, which are associated with the eigenvalues, the partition function *Z* and the free energy *f* in the same level. Thus, the contributions of the nontrivial topological structures are strong for the critical exponents η and ν. The critical exponent α depends on the specific heat $C = -T\frac{\partial^2 f}{\partial T^2}$ that is the second derivative of the free energy *f* with respect to temperature *T*. The topological contributions to the critical exponent α should be small, but the simulations usually have a large deviation with respect to the exact solution (α = 0). This is caused particularly by a fact that it is hard to distinguish the power law of α < 0.2 and the logarithmic singularity (α = 0) [5]. It is suggested to fit the simulations with the logarithmic function for the specific heat. The topological contributions to the free energy *f* of the 3D Ising model is a variable with respect to the change of temperature *T*, which can be viewed figuratively as a displacement. The topological contributions in the first derivative of the free energy *f* can be viewed as a velocity, while the topological contributions in the second derivative of the free energy *f* can be viewed as an acceleration. The values for the velocity and the acceleration can differ. Therefore, the effect of the topological contributions to the critical behaviors in the 3D Ising model can depend on the first or second derivative of the free energy *f*.

Some controversial results exist in literature for the critical exponents of magnetic materials. At first, different universality classes have been reported in various magnetic materials. Even for different compounds/alloys in a same material system, the critical exponents can be quite different. Second, these experimental data have been catalogued to different classes, based on the approximation values of several models, such as the Ising model, the Heisenberg model, etc., which may be far from the exact solutions and may mislead. To clarify the controversial results, we suggest the following strategies: 1) Carefully performing experimental procedures, for instance, using the good samples with high quality (single crystals are better), keeping the equilibrium conditions during magnetic measurements, recording the experimental data as dense as possible at the critical region, fitting the data as accuracy as possible, etc. 2) Regrouping the universality classes with the guidance of the exact solutions of the 3D Ising model, and also the new thoughts on the 3D Heisenberg model (see the next paragraph).

Finally, we pay a special attention on the Heisenberg model. No exact solution has been reported for the 3D Heisenberg model, since the problem is much more complicated than that for the 3D Ising model. To date, the well-accepted approximation values for the critical exponents of the 3D Heisenberg model are $\alpha$ = -0.115, $\beta$ = 0.3645, $\gamma$ = 1.386, $\delta$ =4.802, $\eta$= 0.033 and $\nu$ = 0.705 [62-64,71]. However, similar to the 3D Ising model, the nontrivial topological structures exist also in the 3D Heisenberg model, because they originate from the contradictory of the 2D character of transfer matrices in quantum statistic mechanism and the 3D character of the spin arrangements in a 3D lattice [11,15]. It is expected that the nontrivial topological structures also contribute to the critical behaviors in the 3D Heisenberg model. It is an interesting topic to investigate the topological parts of the critical exponents of the 3D Heisenberg model. One may calculate the critical exponents of the structures illustrated in Figure 5 of ref. [9] (also Figure 1 in ref. [69]), but using the Heisenberg spins to replace the Ising spins. The results obtained in this approach would consist of two kinds of contributions, i.e., local spin alignments and nonlocal long-range spin entanglements (being the topological parts). The topological parts can be evaluated by the difference between these results and the approximation values obtained in the conventional approximation procedures. It would be very significant to catalog the universalities of the 3D Ising class and the 3D Heisenberg class for the critical behaviors in the magnetic materials, based on the calculations with the guidance of the topological contributions.

## 4.Conclusions

In conclusion, this article briefly reviews recent advances in the experiments of the critical exponents in magnetic materials, such as transition-metal intermetallics, Rare-earth transition-metal compounds and manganites. The experimental data confirm the existence of the 3D Ising universality class with the critical exponents $\beta = 3/8$, $\gamma = 5/4$ and $\delta = 13/3$ in the 3D Ising magnets, which affirm the validity of the exact solutions of the 3D Ising models [5]. The topological contributions to the critical behaviors in the 3D Ising model are determined by the difference between the exact solutions and the approximations values. The present work would provide some new insights on the critical behaviors in the 3D Ising magnets and also some implications on the critical behaviors in the 3D Heisenberg model.

**Acknowledgements**

This work has been supported by the National Natural Science Foundation of China under grant number 52031014.

**Data availability** Data available upon request from the author.

**Conflict of interest** The author declares that this contribution is no conflict of interest.